\documentclass[12pt]{article}
\usepackage{latexsym}
\usepackage{amsmath}
\usepackage{wasysym}
\usepackage{amsfonts}
\usepackage{dsfont}
\usepackage{pifont}
\usepackage{bm}
\usepackage{epsf}
\usepackage{epsfig,graphicx}
\usepackage{amsmath,amssymb,bm,epsf,epsfig,graphicx}

\usepackage[mathscr]{eucal}
\usepackage[all]{xy}

\usepackage[vcentermath]{youngtab}

\usepackage[colorlinks=true,linkcolor=blue,citecolor=blue,urlcolor=black,filecolor=black,linktocpage=true]{hyperref}

 \csname
@addtoreset\endcsname{equation}{section}

\def\pe{\prime}
\def\3s{{s \choose 3}}
\def\4s{{s \choose 4}}
\def\5s{{s \choose 5}}
\def\6s{{s \choose 6}}

\def\12{\frac{1}{2}}
\def\fr{\frac}
\def\ft{\footnote}

\def\pr{\partial}

\def\prd{\partial \cdot}

\def\be{\begin{equation}}
\def\ee{\end{equation}}
\def\bea{\begin{eqnarray}}
\def\eea{\end{eqnarray}}
\def\ba{\begin{array}}
\def\ea{\end{array}}
\def\bec{\begin{center}}
\def\ec{\end{center}}

\def\a{\alpha}
\def\b{\beta}
\def\g{\gamma}

\def\d{\delta}

\def\e{\epsilon}
\def\ve{\varepsilon}

\def\h{\eta}

\def\l{\lambda}
\def\L{\Lambda}
\def\m{\mu}
\def\n{\nu}

\def\r{\rho}

\def\vf{\varphi}

\def\o{\omega}

\def\cL{{\cal L}}

\usepackage{color}

\def\tbl{\textcolor{blue}}

\begin{document}

\begin{center}

%%%%%%%%%%%%%%%%%%%%%%%%%%%%%%%%%%%%%%%%%%%%%%%%%%%%%%%%%%%%%%%%%%%%

{\large\sc \bf {On the gauge symmetries of Maxwell-like higher-spin Lagrangians}}\\

%%%%%%%%%%%%%%%%%%%%%%%%%%%%%%%%%%%%%%%%%%%%%%%%%%%%%%%%%%%%%%%%%%%%

\vspace{30pt} {\sc Dario Francia${}^{\; a}$, Simon L.
Lyakhovich${}^{\; \,b}$ and Alexey A. Sharapov${}^{\; \,b}$} \\
\vspace{10pt}
{${}^a$Scuola Normale Superiore and INFN \\ Piazza dei Cavalieri 7, I-56126 Pisa, Italy} \\
e-mail:
{\small \it dario.francia@sns.it }
\vspace{10pt}

{${}^b$\sl\small Department of Quantum Field Theory \\
Tomsk State University, \\
36 Lenin Ave., Tomsk 634050, Russia} \\
e-mail:
{\small \it sll@phys.tsu.ru, \, sharapov@phys.tsu.ru}

%%%%%%%%%%%%%%%%%%%%%%%%%%%%%%%%%%%%%%%%%%%%%%%%%%%%%%%%%%%%%%%%%%%%

\vspace{30pt} {\sc\large Abstract}\end{center}
 In their simplest form, metric-like Lagrangians for higher-spin massless fields are usually assumed to display constrained gauge symmetries, unless auxiliary fields are introduced or locality is foregone. Specifically, in its standard incarnation, gauge invariance of Maxwell-like Lagrangians relies on parameters with vanishing divergence. We find an alternative form of the corresponding local symmetry involving unconstrained gauge parameters of mixed-symmetry type, described by rectangular two-row Young diagrams and entering high-derivative gauge transformations. The resulting gauge algebra appears to be reducible and we display the full pattern of gauge-for-gauge parameters, testing its correctness via the corresponding counting of degrees of freedom. The algebraic techniques applied in this work also allow us to elucidate some general properties of linear gauge systems. In particular, we establish the general fact that any linear local field theory always admits unconstrained,  local, and finitely reducible parameterization of the gauge symmetry. Incidentally, this shows that massless higher spins admit a local  unconstrained formulation with no need for auxiliary fields.

\vfill
\setcounter{page}{1}

\pagebreak

\tableofcontents

%\newpage

%%%%%%%%%%%%%%%%%%%%%%%%%%%%%%%%%%%%%%%%%%%%%%%%%%%%%%%%%%%%%%%%%%%%%

\section{Introduction}\label{sec:intro}

 The goal of this work is to analyze from a novel perspective the gauge symmetry of the higher-spin equations proposed in \cite{SV, CF}. In order to better frame our results it might be useful to first recall some basic facts about the description of massless higher spins.

 Free massless particles of spin $s$ can be described covariantly in terms of rank$-s$ symmetric tensors by means of the following set of conditions
\cite{fierz}\ft{For our notation and conventions the reader can consult \cite{D07}. Here $\vf$ is a rank-$s$ symmetric tensor, whose indices are omitted for simplicity.
Products of different tensors are symmetrised with the minimal number of terms and with no weight factors.  $\vf^{\, \pe}$ denotes the Lorentz trace of $\vf$  while $\prd \vf$  stands for its divergence. Elsewhere in this paper we shall display indices, sometimes according to an alternative compact convention, as in section \ref{sec:solve_traceful}.}:
\be\label{fierz}
\Box  \, \vf \, = \, 0 \, ,\qquad
\prd \vf \, = \, 0 \, ,  \qquad
\vf^{\, \pe} \, = \, 0 \, ,
\ee
while additional non-physical polarisations still contained in $\vf$ can be removed by the on-shell gauge transformation $\d\, \vf = \pr \, \L$, with the parameter $\L$ also satisfying
\be\label{fierzL}
\Box  \, \L \, = \, 0 \, ,\qquad
\prd \L \, = \, 0 \, ,  \qquad
\L^{\, \pe} \, = \, 0 \, .
\ee
 The problem of finding a Lagrangian leading to \eqref{fierz} was first solved long ago by Fronsdal \cite{fronsdal}. As of now,  several
Lagrangian formulations  are known, all generalizing \cite{fronsdal} from various perspectives\ft{ One issue concerns the possibility of removing all the subsidiary conditions \eqref{fierz} and \eqref{fierzL} at the Lagrangian level. In Fronsdal's setting in particular one keeps the trace condition on the gauge parameter, $\L^{\, \pe} = 0$, off-shell. Alternatives to this constraint (and to the double-trace constraint on the gauge field, needed in order to obtain a proper action principle) either exploit auxiliary fields in various forms (see e.g. \cite{PT} for early attempts along these lines, \cite{fs3} for the minimal local extension of Fronsdal's theory, \cite{Maxim} and references therein for the parent approach.)  or they make use of high-spin curvatures, and consequently involve non-local operators \cite{fs1, fms1, D10}.}. A more recent  alternative description of free massless fields \cite{SV, CF} will be the object of the present work.  With respect to \cite{fronsdal} and to its generalizations, the ``Maxwell-like'' Lagrangians of \cite{SV, CF} display a rather peculiar feature: their gauge invariance obtains only if the parameter $\L$ entering the transformation $\d\, \vf = \pr \, \L$ satisfies the condition
\be \label{tdiff}
\partial \cdot \L=0.
\ee
The main observation of  this work is that the transversality condition \eqref{tdiff} can always be solved in terms of fully unconstrained parameters, while keeping the Lagrangians of \cite{SV, CF} unchanged. This shows that the presence of  constrained gauge invariance does not represent an intrinsic feature of Maxwell-like models but rather a choice of parametrization of the corresponding local symmetry. From a more general perspective, and differently from what is usually assumed, this also shows that massless higher spins admit a fully unconstrained local description with no need for additional auxiliary fields.

In order to recall the general setting, let us consider the Lagrangians proposed in  \cite{SV, CF},
\be \label{Mlagr}
\cL \, = \, \12 \, \vf \, M \, \vf \, ,
\ee
where $M$ is the Maxwell operator

\be \label{M}
M \, = \, \Box \, - \, \pr \, \prd \, .
\ee
Under $\d\, \vf = \pr \, \L$ one obtains  $\d \cL \, \sim \, \vf \, \pr^{\, 2} \prd \L$, thus leading to impose on the gauge parameter the transversality condition \eqref{tdiff}.

Further restrictions on the traces of $\vf$ and of $\L$ may or may not be assumed, and would influence the form of the spectrum: for traceless fields and parameters the Lagrangian \eqref{Mlagr} describes the free propagation of a single massless, spin$-s$ particle \cite{SV}, while when traces are kept one obtains a reducible spectrum of massless particles with decreasing spins: $s, \, s-2, \, s-4, \, \cdots $ \cite{CF}. The same spectrum  arises in the symmetric sector of free open string field theory in the tensionless limit \cite{triplets1, triplets2, fs2, st}, to which \eqref{Mlagr} can indeed be related upon partial gauge fixing. Generalizations of \eqref{Mlagr} to the case of mixed-symmetry fields and to (A)dS backgrounds have been studied in \cite{CF}.

The analysis of \cite{SV, CF} shows that the transversality condition \eqref{tdiff} does not hamper the construction of consistent free models. However, the quantization of these models as well as the construction of cubic and possibly higher order vertices have not been attempted yet, and one should expect that the differential constraint \eqref{tdiff} might present a problem at these further stages. Indeed, to begin with, the standard BV quantization formalism \cite{BV} and even the Faddeev-Popov receipt do not apply to the case of local symmetry with parameters subject to differential conditions\ft{Actually \eqref{tdiff} would impose restrictions on the corresponding ghosts, thus reducing the overall set of ghost degrees of freedom similarly to the case of a reducible gauge symmetry.  However, while for the latter case the appropriate sequence of ghosts for ghosts is well known \cite{BV}, a differential constraint like \eqref{tdiff} would force on the corresponding ghost-for-ghost sequence a different ghost number spectrum. This additional complication, to the best of our knowledge, was not considered so far.}. Moreover, on the side of possible deformations of the Lagrangian \eqref{Mlagr} by inclusion of cubic and higher-order terms, one should notice that Noether's second theorem does not hold straight when the gauge parameters are not arbitrary functions, thus raising an issue concerning the applicability of the standard techniques for perturbative inclusion of interactions. A general comprehensive discussion of these and related aspects can be found in \cite{KLS}.

While at this stage it is difficult to assess the very nature, technical or conceptual, of the difficulties related to the differentially constrained gauge parameters, it seems anyway useful to find a solution circumventing any constraints, while still preserving the form of the Lagrangian.

In order to provide a simple illustration of our general idea let us consider the spin$-2$ case and observe that, according to the Poincar\'e lemma, one can solve \eqref{tdiff} as
\be \label{spin2}
\L_{\, \mu}\, =\, \pr^{\, \a}\, \e^{(0)}_{[\a,\, \m]}\,,
\ee
with $\e^{(0)}_{[\a, \, \m]}$ being an arbitrary $2-$form (here square brackets denote antisymmetry)\ft{Solutions to \eqref{tdiff} of the form \eqref{spin2} were proposed long ago in \cite{pope}, for area-preserving diffeomorphisms of two-dimensional surfaces, and were recently considered in \cite{AHV} in the context of finding the correct Faddeev-Popov rules for $4D$ unimodular gravity. The naive Faddeev-Popov quantization procedure indeed does not apply as far as the diffeomorphism parameters are constrained. The Lagrangian \eqref{Mlagr} for $s=2$ reduces to the linearised version of unimodular gravity, where the determinant of the metric is kept fixed and gauge invariance restricts to volume-preserving diffeomorphisms \cite{uni-ham, uni-new}. Leaving aside the notorious difficulties connected with higher-spin interactions, the possibility to implement a similar programme  for $s >2$ should meet in particular the issue of finding the proper deformation of the transversality condition \eqref{tdiff} beyond the linear level.}. Correspondingly, one can write the gauge transformations for the spin$-2$ massless field in the unconstrained form
\be \label{s2gt}
\d \, \vf_{\m \n} \, =\, \pr^{\, \a}  \left(\pr_{\m}\, \e^{(0)}_{[\a,\, \n]}\, + \, \pr_{\n}\, \e^{(0)}_{[\a,\, \m]} \right) \, .
\ee
There are two features of the parametrization \eqref{spin2} and \eqref{s2gt} that are worth stressing:
\begin{itemize}
 \item the variation of $\vf$ contains more than one derivative of the new gauge parameter;
 \item the gauge algebra is {\it reducible}, the pattern of gauge-for-gauge transformations being provided by successive divergences of forms of increasing degrees,
\be \label{spin2gfg}
\begin{split}
&\d\, \e^{(0)}_{[\m, \, \n]} \, = \, \pr^{\, \a}\, \e^{(1)}_{[\a,\, \m, \, \n]} \, , \\
&\d\, \e^{(1)}_{[\m,\, \n, \, \r]} \, = \, \pr^{\, \a}\, \e^{(2)}_{[\a,\, \m, \, \n, \, \r]} \, , \\
&\cdots \, , \\
\end{split}
\ee
where the sequence extends up to the maximal admissible form degree
in a given space-time dimension, i.e. up to the  $D-$form in dimension $D$.
\end{itemize}
From the vantage point of describing the gauge symmetries of second-order Lagrangians for symmetric gauge potentials both features appear unconventional\ft{Reducibility of the gauge symmetry is typical for theories involving tensors with mixed-symmetry \cite{LM}. See \cite{CF} for the corresponding generalization of \eqref{Mlagr}.}. However, they provide the clue to the elimination of the transversality constraints, and the main issue at stake in this paper is to display explicitly the generalization of \eqref{s2gt} to the case of arbitrary spins.

Both for traceful and traceless gauge parameters, whenever $D \geq 4$, the general solution to \eqref{tdiff}  is
\be \label{solve0}
\L_{\, \m_1 \, \cdots \m_{s-1}} \, = \, \pr^{\, \a_1} \, \cdots \, \pr^{\, \a_{s-1}}\, \e^{\, (0)}_{\, \a_1 \, \cdots \a_{s-1}, \, \m_1 \, \cdots \m_{s-1}}\, ,
\ee
with $\e^{(0)}$ taking value in the irrep of $GL(D)$ or of $O(D)$ corresponding to a rectangular tableau with two rows,
\be \label{epsilon}
\e^{(0)}_{\, \a_1 \, \cdots \, \a_{s-1}, \, \m_1 \, \cdots\, \m_{s-1}}:\, \hskip .2cm {\small
\overbrace{\young(\hfil \hfil \cdots \hfil,\hfil \hfil \cdots \hfil)}^{s-1} } \, .
\ee
Here and in the following the representation is constructed so as to leave manifest symmetry along the rows of the corresponding Young diagrams.

Eqs. \eqref{solve0} and \eqref{epsilon} represent the first step in  our analysis, providing the unconstrained form of the gauge symmetry of Lagrangian \eqref{Mlagr}. We first discuss in section \ref{sec:traceful} the multi-particle case, corresponding to traceful fields and parameters, for which the analysis is simpler and whose outcome we use to investigate the traceless case in section \ref{sec:traceless}. Taking into account the full pattern of gauge-for-gauge transformations, that we also display in the corresponding sections, it is then possible to check the consistency of the procedure by counting the degrees of freedom of the system via the general formula \eqref{DoF}, about which we provide some details in section \ref{sec:dof}.

For the spin-2 case the parameter is a divergenceless vector whose dual $(D-1)-$form is then closed, thus allowing to derive the reducibility pattern  \eqref{spin2gfg} by repeated application of the Poincar\'e lemma, as we already mentioned. However, for higher spins we follow a different route and provide a proof of the completeness of the resolutions that we present in sections \ref{sec:traceful} and \ref{sec:traceless} {\it a posteriori}, counting the  number of degrees of freedom that they eventually predict. In order to motivate our results from the first principles, we also discuss our problem from the general mathematical perspective of searching for the kernel of linear differential operators in the space of differential operators. This is a well-known issue in commutative algebra, usually referred to as a \emph{syzygy} problem \cite{E}, whose formulation we recall in section \ref{sec:syz}.  It may be interesting to observe that, while unusual from the more customary field-theoretical viewpoint, the form of the unconstrained gauge parameter \eqref{epsilon} provides indeed the ``natural'' form of the gauge symmetry of the differential operator \eqref{M}, when studied from this general algebraic perspective. In particular, for any linear system of local field equations, like those emerging from \eqref{Mlagr}, the corresponding techniques ensure that the resulting symmetry will not be subject to differential constraints\ft{The same techniques also ensure that the overall reducibility order can never exceed the space-time dimension in any linear theory. In this sense, the possibility of infinite sequences of ghosts for ghosts is a phenomenon attainable only for intrinsically nonlinear systems.}.

In this respect, let us also mention that the mere existence of high-derivative, reducible parametrizations of the gauge symmetry of linear equations is not exclusive of theories with constrained gauge invariance, like \eqref{tdiff}. For instance, maybe surprisingly at first glance, even for the ordinary Maxwell or Fierz-Pauli equations one can ``solve'' for the scalar and the vector parameters in terms of tensor parameters of mixed-symmetry type. In the Outlook we provide some details about this unusual possibility, and conclude our paper displaying consistent patterns of high-derivative reducible transformations for spin-one and spin-two gauge potentials.

%%%%
\section{Multi-particle Lagrangians} \label{sec:traceful}
%%%%

  Our main task in the present work is to reinterpret the gauge invariance of \eqref{Mlagr} under $\d \vf \, = \, \pr \L$, with $\L$  subject to  the transversality condition \eqref{tdiff}, in terms of {\it unconstrained}  gauge parameters, clarifying the features and the meaning of the corresponding picture. We describe first the case of traceful fields and parameters for which the resolution of the constraint is simpler than for the traceless case.

\subsection{Resolution of the transversality constraint} \label{sec:solve_traceful}

As anticipated in the Introduction, the basic idea is to solve for \eqref{tdiff} as follows:
\be \label{solve}
\L_{\, \m_1 \, \cdots \m_{s-1}} \, = \, \pr^{\, \a_1} \, \cdots \, \pr^{\, \a_{s-1}}\, \e^{\, (0)}_{\, \a_1 \, \cdots \a_{s-1}, \, \m_1 \, \cdots \m_{s-1}}\, ,
\ee
where $\e^{(0)}$ takes value in the irrep of $GL(D)$ corresponding to a rectangular tableau with two rows,
\be \label{epsilon0}
\e^{(0)}_{\, \a_1 \, \cdots \, \a_{s-1}, \, \m_1 \, \cdots\, \m_{s-1}}:\, \hskip .3cm
{\small \overbrace{\young(\hfil \hfil \cdots \hfil,\hfil \hfil \cdots \hfil)}^{s-1} } \, ,
\ee
so that the transversality condition \eqref{tdiff} is identically satisfied.

In order to simplify our expressions in the following we shall often denote indices in a symmetrised group with the same symbol, while specifying their overall number by an additional label, e.g.
\begin{align}
&\e^{(0)}_{\, \a_1 \, \cdots \a_{s-1}, \, \m_1 \, \cdots \m_{s-1}} & &\longrightarrow & &
\e^{(0)}_{\, \a_{s-1}, \,  \m_{s-1}} \, ,& & \\
& \pr^{\, \a_1} \, \cdots \, \pr^{\, \a_{s-1}}\, \e^{\, (0)}_{\, \a_1 \, \cdots \a_{s-1}, \, \m_1 \, \cdots \m_{s-1}}\,
& & \longrightarrow  & & (\prd)^{\, \a_{s-1}}\, \e^{\, (0)}_{\, \a_{s-1}, \,  \m_{s-1}}\, . & &
\end{align}
For instance, the gauge transformation of $\vf$ in this notation looks
\be \label{deltaphi}
\d \,  \vf_{\, \m_{s}} \, = \, \pr_{\m} \, \L_{\, \m_{s-1}} \, = \, \pr_{\m} \,  (\prd)^{\, \a_{s-1}}\, \e^{\, (0)}_{\, \a_{s-1}, \,  \m_{s-1}}\, .
\ee
The solution \eqref{solve} to the constraint \eqref{tdiff} features the following two properties:
\begin{itemize}
 \item it displays high derivatives, although referring to a second-order Lagrangian;
 \item it defines a {\it reducible} gauge transformation.
\end{itemize}
Indeed,  under the following variation of $ \e^{\, (0)}$,
\be \label{gfg0}
\d \, \e^{\, (0)}_{\, \a_{s-1}, \, \m_{s - 1}} \, = \,
\pr^{\, \b} \, \e^{\, (1)}_{\, \b, \, \a_{s-1}, \, \m_{s - 1}}\, ,
\ee
with $\e^{\, (1)}$ described by the diagram
\be
\e^{\, (1)}_{\,  \b, \, \a_{s-1}, \, \m_{s - 1}}:\, \hskip .3cm
{\small \overbrace{\young(\hfil \hfil \cdots \hfil,\hfil \hfil \cdots \hfil,\hfil)}^{s-1}} \, ,
\ee
the gauge potential $\vf$ does not change, due to the symmetrization of the index $\b$ with the whole row of $\a$ indices enforced  by the corresponding $s$ divergences. More generally, one can identify a whole chain of reducibility conditions defined via the transformations
\be \label{gfg}
\d \, \e^{\, (k)}_{\, \b^{(k)}, \, \cdots,  \, \b^{(1)}, \,  \a_{s-1}, \,  \m_{s - 1}} \, = \,
\pr^{\, \b^{(k+1)}} \, \e^{\, (k+1)}_{\, \b^{(k+1)}, \, \b^{(k)}, \, \cdots,  \, \b^{(1)}, \,  \a_{s-1}, \,  \m_{s - 1}}\, ,
\ee
where the parameters $ \e^{\, (k)}$ possess the structure of the $GL(D)-$tableaux
\be \label{epsilonk}
\e^{\, (k)}_{\, \b^{(k)}, \, \cdots,  \, \b^{(1)}, \,  \a_{s-1}, \,  \m_{s - 1}}:\, \hskip .3cm
{\small \overbrace{\young(\hfil \hfil \cdots \hfil,\hfil \hfil \cdots \hfil,1,\vdots,k)}^{s-1}} \, , \\
\ee
with $k = 0, \, \cdots, \, D-2$, so that the transformations \eqref{gfg} at level $k$ do not affect the parameter at the $(k-1)-$th generation, i.e.
\be
\d^{\, 2} \, \e^{\, (k-1)}_{\, \b^{(k-1)}, \, \cdots,  \, \b^{(1)}, \, \a_{s-1}, \,  \m_{s - 1}} \, = \, \pr^{\, \b^{(k+1)}} \, \pr^{\, \b^{(k)}} \, \e^{\, (k+1)}_{\, \b^{(k+1)}, \, \b^{(k)}, \, \b^{(k-1)}, \, \cdots,  \, \b^{(1)}, \,  \a_{s-1}, \,  \m_{s - 1}}\, \equiv \, 0 \, .
\ee
In order to prove that \eqref{solve} and \eqref{gfg} provide a resolution for the gauge symmetry of \eqref{Mlagr}, in the next section we perform the degrees of freedom count exploiting the parameters $\e^{(k)}$, checking that it produces indeed the correct result.

In section \ref{sec:syz} we shall discuss how to interpret our results from the algebraic perspective of syzygies \cite{E}. However, let us mention at this point that our solution \eqref{solve} to the transversality condition \eqref{tdiff}, as well as the first reducibility transformation (\ref{gfg0}), also admit a nice explanation in terms of $s$-complexes associated with maximal irreducible tensors\footnote{Given a positive integer $s$,  a $GL$-irreducible tensor is  called maximal if its Young tableau contains no more than $s-1$ columns and at most one row of length $l <s-1$.}. More precisely, the equations $\partial\cdot \Lambda=0$ and $(\prd)^{\, \a_{s-1}}\, \e^{\, (0)}_{\, \a_{s-1}, \,  \m_{s-1}}=0$ can be interpreted as the cocycle conditions for the cochains $\Lambda$ and $\epsilon^{(0)}$. The generalized Poincar\'e lemma  then states that these cocycles are trivial, i.e., coboundaries given by the r.h.s. of (\ref{solve}) and (\ref{gfg0})  \cite{D-VH, BB}. In order to extend the analysis to the higher-order reducibility parameters, that do not meet the maximality condition, one can resort for instance to the prescriptions that were proposed  by Labastida and Morris \cite{LM} in order to describe the gauge  transformations  for massless fields of mixed-symmetry type. Namely, applying the Labastida-Morris rules on the duals of the gauge parameters $\e^{(k)}$ at the various stages (with dualization enforced on each column) one can indeed reproduce, upon dualizing back, the full chain of gauge-for-gauge transformations \eqref{gfg}.

%%%%
\subsection{Degrees of freedom count} \label{sec:traceful_dof}
%%%%

In the absence of second-class constraints\ft{As it is the case for the Lagrangian \eqref{Mlagr}. See \cite{CF} for an explicit discussion of this point.}, the number $N$ of degrees of freedom described by second-order Lagrangian field equations is provided by the formula \cite{KLS}
\be \label{DoF}
N \, = \, n \, + \, \sum_{k, m=0}\, (-1)^{k+1}\, (m\, +\, 1)\, r_m^{(k)} \, ,
\ee
where $n$ is the number of field components, while $r_m^{(k)}$ is the number of gauge generators at the $k$-th reducibility generation, of the $m-$th differential order. For instance, with reference to the pattern of gauge-for-gauge transformations proposed in the previous section, the contribution from $\e^{(0)}$ would be encoded in $r_s^{(0)}$, the dimension of the tableau \eqref{epsilon0} appearing in $\d \vf$ with a total of $s$ derivatives. More generally, the parameters  $\e^{(k)}$ would be accounted for in the numbers $r_{s+k}^{(k)}$, dimensions of the tableaux \eqref{epsilonk} contributing to $\d \vf$ with a total of $s + k$ derivatives. See section \ref{sec:dof} for some basics on the covariant counting of degrees of freedom. Here we would like to apply \eqref{DoF} to the resolution described in the previous section and check the resulting number of degrees of freedom against the corresponding result found in \cite{CF}, in the analysis exploiting the gauge parameter $\L$ subject to \eqref{tdiff}.

Let us mention that the nature of the resulting degrees of freedom was also discussed in \cite{CF}, where it was shown how to write the Lagrangian \eqref{Mlagr} in a diagonal basis so as to explicitly identify its particle content, beyond the overall number of degrees of freedom. Moreover, the diagonal basis also makes it manifest that all the kinetic terms in the corresponding single-particle Lagrangians carry the same sign, and thus the theory does not propagate ghosts.

In order to apply \eqref{DoF} we need the dimensions of the diagrams corresponding to the parameters $\e^{(k)}$ in \eqref{epsilonk}, appearing in $\d \vf$ with a total of $s + k$ derivatives, and thus contributing to \eqref{DoF} with a weight equal to $s + k + 1$; making use of the formulae given in section \ref{sec:dimension} we obtain
\be \label{dimeps}
\mbox{dim}_{GL(D)} \, {\small \overbrace{\young(\hfil \hfil \cdots \hfil,\hfil \hfil \cdots \hfil,1,\vdots,k)}^{s-1}}  \, = \, \fr{s-1}{(s + k)\, (s + k -1)} \, {D - 2 \choose k} \, {D + s - 3 \choose s - 1} \, {D + s - 2 \choose s - 1} \, ,  %\hskip.3cm o(\pr^{s + k})
\ee
with $k = 0, \cdots, D-2$.

As a first test we  compute the number of degrees of freedom when $\vf$ is a rank-$s$ tensor in $D = 4$; the relevant fields and their dimensions are given by
\begin{align}
& \vf^{\, \ \  }: && {\small \overbrace{\young(\hfil \hfil \cdots \hfil \hfil)}^{s}} \, , &&  {s + 3 \choose s}    \\
&  \e^{\, (0)} : && {\small \overbrace{\young(\hfil \hfil \cdots \hfil,\hfil \hfil \cdots \hfil)}^{s-1}}\, , && \fr{1}{s} {s + 1 \choose s-1} {s+2 \choose s-1},  && o(\pr^{s})\\
&  \e^{\, (1)} : && {\small \overbrace{\young(\hfil \hfil \cdots \hfil,\hfil \hfil \cdots \hfil,\hfil)}^{s-1}}\, , && 2 \fr{s-1}{s (s + 1)} {s + 1 \choose s-1} {s+2 \choose s-1}, && o(\pr^{s+1})     \\
&  \e^{\, (2)} : && {\small \overbrace{\young(\hfil \hfil \cdots \hfil,\hfil \hfil \cdots \hfil,\hfil,\hfil)}^{s-1}} \, , && \fr{s-1}{(s+1) (s + 2)} {s + 1 \choose s-1} {s+2 \choose s-1},  && o(\pr^{s+2})\, ,
\end{align}
where we also kept track of the overall order of derivatives for each gauge parameter.
Upon substitution in \eqref{DoF} one finds
\be
{s + 3 \choose s} -  {s + 1 \choose s-1} {s+2 \choose s-1}\, \left[ \fr{s+1}{s}  - 2  \fr{(s-1) (s+2)}{s  (s+1)}  +
\fr{(s-1) (s+3)}{(s+1) (s + 2)}   \right]  =  s  +  1 \, ,
\ee
matching the dimension of a symmetric, rank-$s$ tensor of $GL(2)$, and thus describing correctly the degrees of freedom of the chain of massless particles of spin $s, s-2, s-4, \cdots $ that constitute indeed the spectrum of \eqref{Mlagr} in the absence of trace conditions.

By an inductive argument, one can then prove the following relation valid in arbitrary $D$:
\be
\begin{split}
&{D + s - 1 \choose s} -  {D + s - 3 \choose s-1} {D + s -2 \choose s-1}  \sum_{k = 0}^{D-2} (-1)^k  \fr{(s-1) (s+k + 1)}{(s+k) (s+k-1)}
{D - 2 \choose k} \\
& = \,{D + s - 3 \choose s} ,
\end{split}
\ee
thus confirming  that the physical polarisations associated to the equations $M \vf = 0$, with unconstrained gauge invariance described by the pattern \eqref{deltaphi} and \eqref{gfg},  match those of the corresponding irreducible representation of $GL(D-2)$.

%%%%
\section{Single-particle Lagrangians} \label{sec:traceless}
%%%%

The same Lagrangian \eqref{Mlagr} describes the propagation of a single, massless particle of spin $s$ if, in addition to the
transversality condition  \eqref{tdiff} for the gauge parameter $\L$, one also requires  both
the field and the parameter to be traceless \cite{SV}:
\be \label{traces}
\begin{split}
& \vf^{\, \pe} \, = \, 0 \, ,      \\
& \L^{\, \pe} \, = \, 0 \, .
\end{split}
\ee
In this section, we would thus like to discuss the resolution of the transversality constraint \eqref{tdiff} under the further restrictions \eqref{traces}.

\subsection{Resolution of the transversality constraint} \label{sec:solve_traceless}

In the simplest attempt to solve the system $\prd \L = 0$ and $\L^{\, \pe} = 0$ one would try the resolution found for the traceful case  supplemented by the condition that the parameters $\e^{\, (k)}$  in \eqref{epsilonk} take values in the corresponding irreps of $O(D)$. However, this identification cannot apply to the last two parameters of the pattern, $\e^{\, (D-3)}$ and $\e^{\, (D-2)}$, whose tableaux do not satisfy the condition
\be
n_1 \, + n_2 \, \leq \, D \, ,
\ee
where $n_1$ and $n_2$ denote the lengths of the first two columns, and thus do not exist as representations of $O(D)$\ft{See e.g. \cite{hamermesh} \textsection  $10.6$.}.

As a  concrete instance of this issue let us consider the spin$-3$ case in $D=4$, where one can still solve the rank$-2$ traceless parameter as
in \eqref{solve}
\be \label{solve3}
\L_{\, \m \m} \, = \, \pr^{\, \a} \, \pr^{\, \a} \, \e^{\, (0)}_{\, \a \a, \, \m \m}\, ,
\ee
with $\e^{\, (0)}_{\, \a \a, \, \m \m}$ to be now interpreted as a two-row, window diagram in $O(D)$. The difficulty arises when  evaluating the gauge-for-gauge invariance associated to \eqref{solve3}, since the corresponding solution of the traceful case, schematically given by
\be
\d \, \e^{\, (0)} \, = \, \pr^{\, \g} \, {\small \young(\hfil \hfil,\hfil \hfil,\g)} \, ,
\ee
does not involve an admissible representation of $O(4)$. One is thus led to consider the possibility that for this class of tableaux the pattern of reducibility of the unconstrained gauge transformation involve more derivatives than for the traceful case, so as to compensate for the indices that cannot be carried by the representation itself. In particular, in the example under consideration, the minimal modification consists in removing one index from the corresponding diagram and trying the schematic ansatz
\be \label{teps1}
\d \, \e^{\, (0)} \, \sim \, \pr^{\, \g} \, {\small \young(\hfil \hfil,\hfil \pr,\g)}\, ,
\ee
where the symbol of the gradient $\pr$ in the diagram pictorially denotes an $O(D)-$projection of the derivative of the diagram without the corresponding box; i.e.
\be
{\small \young(\hfil \hfil,\hfil \pr,\hfil)} \, \equiv \, Y^{}_{\tiny \young(\hfil \hfil,\hfil \hfil,\hfil)} \, \left\{\pr \, \small \young(\hfil \hfil,\hfil,\hfil)\right\} \, .
\ee
Thus, the parameter $\tilde{\e}^{\, (1)}$ implicitly defined in \eqref{teps1} has the symmetry of a $\{2, \, 1, \, 1\}-$tableau in $O(4)$,
with the missing box needed to match the indices of $\e^{\,(0)}$ essentially substituted by one additional gradient. More explicitly,
\be\label{ww}
\d \, \e^{\, (0)}_{\, \a \a, \,  \m \m} \, = \, \pr_{\a}\, \pr^{\, \g} \, \tilde{\e}^{\, (1)}_{\g, \, \a, \, \m \m} \, - \,
 \pr_{\m}\, \pr^{\, \g} \, \tilde{\e}^{\, (1)}_{\g, \, \a, \, \a \m} \, .
\ee
Again, the transformations (\ref{ww}) are reducible and $\e^{(0)}$ does not vary when the tensor $\tilde{\e}^{\, (1)}$ transforms as follows
\be\label{ww2}
\d \, \tilde{\e}^{\, (1)}_{\, \g, \a, \,  \m \m} \, = \, \pr_{\m}\, \tilde{\e}^{\, (2)}_{\, \g,\, \a, \, \m} \, - \,
\fr{1}{D - 2} \, \left(2 \, \h_{\, \m \m} \, \pr^{\, \r} \, \tilde{\e}^{\, (2)}_{\, \g,\, \a, \, \r} \, + \,
\h_{\, \m \a} \, \pr^{\, \r} \, \tilde{\e}^{\, (2)}_{\, \g,\, \r, \, \m} \, + \,
\h_{\, \m \g} \, \pr^{\, \r} \, \tilde{\e}^{\, (2)}_{\, \r,\, \a, \, \m}  \right) \, ,
\ee
where $ \tilde{\e}^{\, (2)}$ is a $3-$form. The last transformation is already irreducible and indeed counting the physical degrees of freedom by means of \eqref{DoF} we find
$$
N\, =\, 16\, -\, 4\cdot 10 \,  +\, 6\cdot 9\, -7\cdot 4\, =\, 2\,.
$$
To complete the analysis of the spin$-3$ case in $D=4$, let us also notice that, via dualization, it is possible to represent $\tilde{\e}^{(1)}$ and $\tilde{\e}^{(2)}$ in a simpler form, as a traceless rank$-2$ tensor and a rank$-1$ tensor, respectively,
\begin{align}
&& &\tilde{\e}^{\, (1)} \, : \, {\small \young(\hfil \hfil,\hfil,\hfil)} & & \longleftrightarrow  &
&\e^{(1)} \, : \, {\small \young(\hfil \hfil)}\, , & &&\\
&& &\tilde{\e}^{\, (2)}\, : \, {\small \young(\hfil,\hfil,\hfil)} \,
&  &\longleftrightarrow &  & \e^{\, (2)}\, : \, {\small \young(\hfil)}\, , & &&
\end{align}
so that the final unconstrained pattern of gauge transformations can be presented in the following form:
\begin{align}
&& &  \e^{\, (0)} : && {\small \young(\hfil \hfil,\hfil \hfil)}\, ,  && o(\pr^{3})\, , &&\\
&& &  \e^{\, (1)} : && {\small \young(\hfil \hfil)}\, ,  && o(\pr^{5}) \, ,    && \\
&& &  \e^{\, (2)} : && {\small \young(\hfil)} \, , && o(\pr^{6})\, .  &&
\end{align}

In section \ref{sec:spinor} we discuss the spin$-s$ case in $D=4$ in spinorial notation. At any rate, the example of spin $3$ already conveys the essential difference between the multi-particle and the single-particle cases, while also providing the source of inspiration for our general ansatz. The latter can be described as follows:
\begin{itemize}
 \item For $D\geq 4$\ft{In section \ref{sec:spin3d3} we discuss explicitly the case $D = 3$.} we solve the transversality condition for transverse-traceless parameters as in \eqref{solve}, and assume the reducibility pattern to be given by \eqref{epsilonk} up to $k \leq D-4$, with the proviso that all diagrams are to be interpreted as taking values in $O(D)$.
 \item Since $\e^{\, (D-3)}$ as defined in \eqref{epsilonk} does not exist in $O(D)$, for the gauge-for-gauge invariance at the level $D-3$ we consider, schematically,
\be \label{eps4}
\d \, \e^{\, (D-4)} \, \sim \, \pr^{\, \b} \,
{\small \overbrace{\young(\hfil \hfil \cdots \hfil,\hfil \pr \cdots \pr,1,\vdots,\b)}^{s-1}} \, , \\
\ee
where $\b = D-3 $; more precisely
\be \label{proj}
\d \, \e^{\, (D-4)}_{\, \b^{(D-4)}, \, \cdots,  \, \b^{(1)}, \, \a_{s-1}, \,  \m_{s - 1}} \, = \,
Y^{O(D)}_{\{s-1, \, s-1, \, 1, \, \cdots, \, 1\}}
\pr_{\, \a_{s-2}}\, \pr^{\, \b^{(D-3)}} \,  \tilde{\e}^{\, (D-3)}_{\, \b^{(D-3)}, \, \b^{(D-4)}, \, \cdots,  \, \b^{(1)}, \,  \a, \,  \m_{s - 1}}\, ,
\ee
where $Y^{O(D)}_{\{s-1, \, s-1, \, 1, \, \cdots, \, 1\}}$ defines the projector onto the irrep matching the structure of $\e^{(D-4)}$. The parameter $\tilde{\e}^{\, (D-3)}$ in \eqref{eps4} is $O(D)-$irreducible and can be dualised to a rank$-(s-1)$ symmetric and traceless tensor, that we shall call once again $\e^{\, (D-3)}$ to avoid introducing new symbols:
\be
\tilde{\e}^{\, (D-3)} \hskip .5cm \longleftrightarrow \hskip .5cm  \e^{\, (D-3)} \, : \, \overbrace{\young(\hfil\hfil\cdots\hfil)}^{s-1}\, .
\ee
\item The gauge-for-gauge transformation of $\tilde{\e}^{\, (D-3)}$ is schematically identified as follows:
\be \label{eps3}
\d \, \tilde{\e}^{\, (D-3)}_{\, \b^{(D-3)}, \, \cdots,  \, \b^{(1)}, \, \a, \,  \m_{s - 1}} \, \sim \,
{\small \overbrace{\young(\hfil \hfil \cdots \pr,\a,1,\vdots,\b)}^{s-1}} ,
\ee
where $\b = D-3$ as above, while the gradient in the last box of the first row pictorially suggests that, upon substitution in \eqref{eps4}, the corresponding variation of $\e^{\, (D-4)}$ would vanish identically. Finally, we notice once again that the parameter $\tilde{\e}^{\, (D-2)}$ implicitly defined in \eqref{eps3} is dual to a symmetric, traceless, rank$-(s-2)$ tensor, which makes it easier to compute its dimension:
\be
\tilde{\e}^{\, (D-2)} \hskip .5cm \longleftrightarrow \hskip .5cm \e^{\, (D-2)} \, : \, \overbrace{\young(\hfil\hfil\cdots\hfil)}^{s-2}\, .
\ee
\end{itemize}

To summarize, our resolution for the transverse and traceless gauge transformation of the single-particle case  is described, for the first $D-3$ parameters, by the same set of tensors $\e^{(k)}$ presented in \eqref{epsilonk} for the traceful case, to be interpreted as tableaux in $O(D)$. The last two parameters in the pattern are different, and after dualization can be identified with traceless symmetric tensors of rank $s-1$ and $s-2$, respectively, contributing to the gauge transformation of $\vf$ with a total number of derivatives that is displayed below:
\begin{align}
&& &  \e^{\, (D-3)} : && {\small \overbrace{\young(\hfil \hfil \cdots \hfil\hfil)}^{s-1}}\, ,  && o(\pr^{2s + D - 5}) \, ,    && \\
&& &  \e^{\, (D-2)} : && {\small \overbrace{\young(\hfil \hfil \cdots \hfil)}^{s-2}} \, , && o(\pr^{2s + D - 4})\, .  &&
\end{align}

The counting of degrees of freedom that we perform in the next section proves the pattern we propose is indeed a resolution of the constrained gauge transformation. The possibility to identify it as  the {\it minimal} resolution requires more general considerations that we postpone to  section \ref{sec:syz}.

\subsection{Degrees of freedom count} \label{sec:traceful_dof}

We would like to test the proposed resolution of the gauge symmetry defined by  \eqref{tdiff} and \eqref{traces}, checking that the corresponding number of degrees of freedom matches the number of polarizations of a massless particle of spin $s$ in $D$ space-time dimensions. To this end we shall need the dimensions of the corresponding tableaux, given by the following formulae (see section \ref{sec:dimension}):
\be \label{dimwindow}
\mbox{dim}_{\, O(D)} \, {\small \overbrace{\young(\hfil \hfil \cdots \hfil,\hfil \hfil \cdots \hfil)}^{s}}  \, = \, \fr{1}{s \, {s + 1\choose 3}} \, {D + s - 4  \choose s - 2} \, {D + s - 5 \choose s - 1} \,  {D + 2 s -2  \choose 3} \, ,
\ee
\be \label{dimepsk}
\begin{split}
\mbox{dim}_{\, O(D)} \, {\small \overbrace{\young(\hfil \hfil \cdots \hfil,\hfil \hfil \cdots \hfil,1,\vdots,k)}^{s-1}}  \, = &  \fr{1}{4\, {2k + 2 \choose k-1} \, {k + 3 \choose k-1}} \,  \fr{{s+k-2 \choose k -1}}{{s + k \choose 2k + 2}}\, { D + s  - k - 6 \choose s - k -2} \, {D - 4 \choose k} \times \\
& { D + s  - 5 \choose k -1}  \, {D + 2 s - 4 \choose 3} \, {D + s - 4 \choose s - 2} \, .
\end{split}
\ee

\noindent
More precisely \eqref{dimepsk} holds for $k \leq \left[\fr{D-4}{2}\right]$, with $[a]$ denoting the integer part of $a$, while for $\fr{D-4}{2} < k \leq D-4 $ the corresponding diagrams possess the same dimension as their dual counterparts in the first set,
\be \label{dualk}
\e^{\, (k >  \fr{D-4}{2})} \, \sim \, \e^{\, (D - k - 4)} \, .
\ee
Finally, the parameters $ \e^{\, (D - 3)}$ and  $\e^{\, (D - 2)}$ are described by symmetric and traceless tensors of rank $s-1$ and $s-2$, respectively, whose dimension can be computed both from the following expression
\be
\mbox{dim}_{\, SO(D)} \, {\small \overbrace{\young(\hfil \hfil \cdots \hfil)}^{\, s}} \, = \, \fr{D + 2s - 2}{D + s - 2} \, {D + s -2 \choose s}\, .
\ee
We are thus in the position to count the degrees of freedom and, according to the general formula \eqref{DoF}, check the equality
\be \label{dof_traceless}
\begin{split}
\vf^{(s)}_{D} \, & - \, \sum_{k = 0}^{[\fr{D-4}{2}]} \, (-1)^k \,
\left[(s + k + 1) \, + \, (-1)^D \, (s + D - k - 3) \right] \, \e^{(k)}\\
& + (-1)^{D} \, \left(s + \fr{D}{2} - 1 \right)\, \d_{[\fr{D-4}{2}], \, [\fr{D-3}{2}]} \,
\e^{([\fr{D-4}{2}])} \\
& + (-1)^D \,  (2s + D - 4) \, \e^{(D-3)} \,+ \, (-1)^{D+1} \, (2s + D - 3) \, \e^{(D-2)} \\
& = \, \vf^{(s)}_{D-2}\, ,
\end{split}
\ee
where for simplicity the symbol of a given tensor stands for the dimension of the corresponding $D-$dimensional tableau, while the r.h.s. of the equality is to be evaluated in $D-2$ dimensions. In the first line we exploited  the duality \eqref{dualk} to pair terms represented by the same tableau, and thus summing the coefficients depending on their differential order in $\d \vf$. The term in the second row is only present for even $D$ and is meant to avoid the double counting of the corresponding self-dual diagram computed from the last term in the previous sum. Finally, in the third row we are summing over the contribution of the last two parameters, $\e^{(D-3)}$ and $\e^{(D-2)}$, whose structure and differential order are specific of the traceless case.

The validity of \eqref{dof_traceless} can be proven by induction,  first checking it for a specific case, e.g. spin $s$ in $D=5$, and then observing that, assuming it to hold for arbitrary $D$,  its validity for $D + 1$ follows by rescaling the corresponding index everywhere.

%%%%
\section{Gauge algebras and syzygies} \label{sec:syz}
%%%%

In this section we shall discuss the general problem of identifying the complete gauge symmetry for local field theories at the level of their free equations, so as to frame the main results of this paper in a more general mathematical setting.

A general system of free equations for the fields $\phi$ can be written schematically as
\be \label{g2l}
A^{a_0}_{a_1}(\partial) \, \phi^{a_1} =0
\, .
\ee
Here we do not assume the field equations  \eqref{g2l} to come from the least action principle, so that the (multi-)indices $a_0$ and $a_1$ labeling  the components of field and equations may be completely unrelated. The entries of the matrix $A^{a_0}_{a_1}$ are polynomials in the partial derivatives $\partial_\mu$. For the Lagrangian \eqref{Mlagr}, the fields $\phi^{a_1}$ correspond to symmetric traceful or traceless tensors in Minkowski space while $A$ is to be identified with the Maxwell operator \eqref{M}. In this case, the multi-indices $a_0$ and $a_1$ coincide, and $A=M$ is a symmetric matrix whose entries are quadratic  in  $\partial_\mu$.

A one-parameter gauge transformation $\delta_\epsilon \phi$  must leave the equations invariant,
\be\label{ggt} \delta_\epsilon ( A^{a_0}_{a_1}\, (\partial) \,
 \phi^{a_1} ) \, = \,  A^{a_0}_{a_1}\, (\partial)\, \delta_\epsilon \, \phi^{a_1} \, \equiv \, 0 \, . \ee
The transformation is supposed to be field-independent and local,
\be \label{ggg}
\delta_\epsilon  \, \phi^{a_1} (x)\, = \, R^{a_1}\,  (\partial ) \, \epsilon (x) \, .
\ee
Locality  means that the gauge generator $R^{a_1} (\partial )$ is polynomial in the partial derivatives $\partial_\mu$. Since the gauge parameter
$\epsilon$ is an arbitrary function of $x$, the gauge generators must satisfy the linear homogeneous equation
\be \label{gi}
A^{a_0}_{a_1}(\partial)\, R^{a_1} \, (\partial) \, = \, 0 \, .
\ee
In this general setting, the problem of identifying the complete gauge symmetry for the field equations (\ref{g2l}) (or for the corresponding Lagrangian, if it exists) reduces to the issue of finding an (over-)complete basis of right null-vectors for the matrix $A$. The null-vectors are assumed to be polynomial in $\partial$, to have a local gauge symmetry. If $R_2=(R_{a_2}^{a_1})$ is such a basis, then any solution to
(\ref{gi}) can be written in the form
\begin{equation}
R^{a_1}=R_{a_2}^{a_1}R^{a_2}
\end{equation}
for some polynomial vector $R^{a_2}(\partial)$. Notice that, as already mentioned, in general the indices $a_1$ and $a_2$ may run over different sets, so that $R_2$ is a rectangular matrix with polynomial entries.

Unlike the case of linear systems over numerical fields like $\mathbb{R}$ and $\mathbb{C}$, where one can aim to find  a basis of linearly independent solutions, in the case of linear equations over the polynomial ring $\mathcal{R}=\mathbb{R}\, [\partial_1,\ldots, \partial_D]$ the problem is more tricky. On the one hand, by the  Hilbert Basis Theorem \cite{E}, it is  always possible to choose  a \textit{finite} number of polynomials such that any solution can be expanded over them with polynomial coefficients. The polynomials spanning the solution space are called the generating  (or complete) set of solutions. On the other hand, the tricky issue is that in certain cases \textit{every} complete set of solutions is linearly dependent over $\mathcal{R}$, or in other words, every  complete set of solutions is actually over-complete and no basis of linearly independent solutions can exist. When this happens, to understand the structure of the solution space one needs to compute the dependency relations among the generators.  This is a well-known issue in commutative algebra usually referred to as a \emph{syzygy problem} \cite{E}. It  provides an algebraic counterpart of the problem of finding the full set of gauge-for-gauge generations in a theory with local symmetry.

The polynomial, right null-vectors of the matrix $A$  form a linear space over $\mathcal{{R}}$, called the \emph{first syzygy module}, having the generators of gauge symmetry as generating set. As mentioned above, there might be linear relations among the generators of the first syzygy module, corresponding to linear dependencies of the gauge symmetries, i.e., there might exist $R^{a_2} (\partial) \neq 0 $ such that $R_{a_2}^{a_1}(\partial )R^{a_2} (\partial)=0$. The whole set of such linear relations form a vector space over $\mathcal{R}$, the space of syzygies for syzygies, also known as the \emph{second syzygy module} (or second syzygy, for short).  From the viewpoint of physics, the existence of a nontrivial second syzygy means that the gauge symmetry is reducible with the generating set in the second syzygy providing generators of gauge-for-gauge transformations. Continuing in this way one obtains  a sequence of matrices $R_k=\{R^{a_{k-1}}_{a_{k}}\, (\pr)\}$, $k=2,3,\ldots$, with polynomial entries such that $R_k R_{k+1}=0$. One can think of this sequence as a proper substitution for the notion of general solution to the system of linear equations over a numerical field. (In the latter case the sequence collapses to a single matrix $R_2=\{R^{a_1}_{a_2}\}$, whose columns define a basis of solutions.)

All that have been said above can be concisely rephrased in the language of commutative algebra.  From the algebraic viewpoint, the right kernel of the matrix $M$ is a finitely generated $\mathcal{R}$-module, which is not free in general, but is a submodule of a free module $\mathcal{R}^{r_1}$. Computing chain of syzygies, one gets the  exact sequence of homomorphisms of free $\mathcal{R}$-modules
\begin{equation}\label{Resol}
\xymatrix{\cdots
\ar[r]^{R_4}&\mathcal{R}^{r_3}\ar[r]^{R_3}&\mathcal{R}^{r_2}\ar[r]^{{R_2}}&\mathcal{R}^{r_1}\ar[r]^{{A}}&\mathcal{R}^{r_0}}\,,
\end{equation}
which provides a \textit{free resolution} of the quotient module $M=\mathcal{R}^{r_0}/\mathrm{Im} A$.  The exactness means that each map is a surjection onto the kernel of the following map and the ranks of the free modules $\mathcal{R}^{r_k}$ are determined by the size of the syzygy matrices $R_k$. In this language, the module of solutions to equations (\ref{gi}) is given by $\mathrm{ker} A$, the module of syzygies of $\mathrm{ker} A$ is $\mathrm{ker} R_2$, the module of second syzygies of $\mathrm{ker} A$ is $\ker R_3$, and so on.

The free resolution (\ref{Resol}) is by no means unique and strongly depends on the choice of generators of the syzygy modules. Different choices may result in resolutions of different length. (Including infinity.) However, the Hilbert Syzygy Theorem ensures the existence of a resolution of length at most $D$, where $D$ is the number of formal variables generating the polynomial ring $\mathcal{R}=\mathbb{R}[\pr_1,\ldots, \pr_D]$.  The minimal possible number of steps before the sequence of syzygies terminates is known as the \emph{global  dimension of the module}. It is a minimal free resolution that defines the gauge structure of the theory (\ref{g2l}). The reducibility order is the global  dimension of the module $M$. It is well known  that the minimal resolution is essentially unique \cite{E}. Among the other things, this proves that \emph{the gauge symmetry of any free  field theory is finitely reducible}. Given (a not necessarily minimal) resolution (\ref{Resol}), one can write the complete gauge symmetry transformation as well as the full chain of gauge-for-gauge transformations:
\begin{equation}
\delta_{\epsilon_2}\phi^{a_1}=R^{a_1}_{a_2}(\partial)\epsilon^{a_2}\,,\qquad \delta_{\epsilon_{k+1}}\epsilon^{a_{k}}=R^{a_k}_{a_{k+1}}(\partial)\epsilon^{a_{k+1}}\,,\quad k=2,3, ...\, .
\end{equation}

The notion of free resolution also applies to the study of gauge (or Noether) identities between the free field equations (\ref{g2l}). These are identified with the field-independent differential operators $L_{a_0}(\partial)$ annihilating the field equations on the left, i.e.,
\begin{equation}
L_{a_0}(A^{a_0}_{a_1}\phi^{a_1})\equiv 0\,.
\end{equation}
Since the $\phi$'s are considered to be arbitrary functions of the $x$'s, this means the following linear equations for the $L$'s:
\begin{equation}
L_{a_0}(\partial)A^{a_0}_{a_1}(\partial)=0\,.
\end{equation}
So, one can identify the generators of gauge identities with the left polynomial null-vectors of the matrix $A$. These vectors form a linear space over $\mathcal{R}$ and to study its structure one should construct a free resolution for the left kernel of the matrix $A$ in much the same way as we did for the right kernel. For Lagrangian equations the matrix of the wave-operator $A$ is square and  symmetric, so that the left and right kernels essentially coincide. As a result the corresponding free resolutions are obtained from each other by formal transposition\footnote{More precisely, we have the following identifications for the indices and generators: $L_{a_{1-k}}^{a_{-k}}(\partial)=R_{a_{k+1}}^{a_{k}}(-\partial)$. The additional ``minus'' is due to integration by parts. This correspondence is a consequence  of the second Noether theorem.}. In the case of general non-Lagrangian equations the free resolutions for the left and right kernels of the rectangular matrix $A$ may be completely different.

Let us stress once more that, even though the original equations are of second order (i.e. the polynomials in $A^{a_0}_{a_1}$ are all quadratic in $\partial$), nothing forces the syzygies to be generated by polynomials of lower order. In this sense, although unfamiliar in the standard approach to gauge theories, there is no real surprise in the fact that second-order Lagrangians can enjoy higher-order gauge symmetries.  As we actually saw in sections \ref{sec:traceful} and \ref{sec:traceless} this is just the case for the Maxwell-like theory \eqref{Mlagr}. In the conventional description of its gauge symmetry this property is somehow hidden by the choice of presenting the gauge transformation in the standard (first-order) form $\d \, \vf  =  \pr \, \L$, subject to the differential constraints \eqref{tdiff}.

The above algebraic technique can be immediately applied to the specific case of interest for us in this work. Notice however  that our strategy in sections \ref{sec:traceful} and \ref{sec:traceless} was somewhat  indirect in the first step: we deduced unconstrained gauge transformations by proposing explicit solutions to the transversality constraints (\ref{tdiff}) without providing in principle a direct proof that we were not missing independent gauge transformations. What made our construction self-contained is that we could exploit the covariant counting of degrees of freedom of \cite{KLS} (see section \ref{sec:dof}) and compare with the independent counting of the degrees of freedom provided in \cite{SV, CF} for the model described by \eqref{Mlagr}. Introducing multi-indices $a_0=(\mu_2,\ldots,\mu_{s-1})$ and $a_1=(\nu_1,\ldots, \nu_{s-1})$ we can write (\ref{tdiff}) in the form (\ref{g2l}) with
\begin{equation}\label{AX}
{\phi}^{a_1}\, (\partial)=\Lambda_{\nu_1\cdots\nu_{s-1}}\,, \qquad
A^{a_0}_{a_1}\, (\partial)\, =\,
\partial^{(\nu_1}\delta^{\nu_2}_{\mu_2}\cdots
\delta^{\nu_{s-1})}_{\mu_{s-1}}\, ,
\end{equation}
where the round brackets denote symmetrization of the corresponding indices. One can then go through the general analysis, looking for the minimal resolution, whose existence is guaranteed by the Hilbert Syzygy Theorem.Our analysis shows that the global dimension of the module $M$ associated with the transversality condition (\ref{tdiff}) reaches the upper bound established by the Hilbert theorem. So, in $D$ dimensions we have the $(D+2)$-term exact sequences
\begin{equation}\label{Res}
\xymatrix{0\ar[r]&\mathcal{R}^{r_D} \ar[r]^{R^D}&
\mathcal{R}^{r_{D-1}}\ar[r]^{R^{D-1}}  &\cdots
\ar[r]^{R^3}&\mathcal{R}^{r_2}\ar[r]^{{R^2}}&\mathcal{R}^{r_1}\ar[r]^{{A}}&\mathcal{R}^{r_0}}\,,
\end{equation}
where the ranks of the free modules depend on the value of the spin $s$ as well as on the algebraic property of the fields (traceful or traceless). The $0$ at the left-end of the exact sequence implies that the module of the $(D-1)-$th syzygies of $\mathrm{ker} A$ is free.

At present, there is a great deal  of computational techniques and software packages for computing  minimal resolutions of polynomial modules.  In performing some checks we used the computer algebra system \textsc{Singular}\footnote{It is freely available for many platforms at the website http://www.singular.uni-kl.de.}. There are, however, two drawbacks with these packages. First, the computational algorithms is such that the output syzygy matrices $R_k$ are not given in  Lorentz covariant form\footnote{One can argue that all the syzygy matrices must be Lorentz covariant whenever the original equations (\ref{AX}) enjoy Lorentz covariance.}. Second, the program works only with a completely specified  set of equations. It is impossible, for example,  to keep $D$ and $s$ as free parameters defining the system and get the solutions for all the cases of interest at once. Still, one can explicitly compute minimal free resolutions for specific cases, e.g.  the pairs $D=3,4,5$ and $s=2,3,4$ and then identify the corresponding free modules $\mathcal{R}^{r_k}$  with the spaces of Young's tableaux of a given shape, while also writing the output syzygy matrices $R_k$ in Lorentz-covariant form. As already mentioned, checking the number of degrees of freedom provides the ultimate test of the overall procedure.

%%%%
\section{Outlook} \label{sec:out}
%%%%

 In this work we showed how to interpret the gauge symmetry of Maxwell-like Lagrangians \eqref{Mlagr} in terms of gauge parameters not subject
to differential conditions like \eqref{tdiff}.  Unlike the constrained parametrization, the resolutions we found should allow a  standard quantization of these systems, while their possible role in the construction of non-linear/non-abelian deformations  of the gauge algebra is to be investigated.
In this respect, however, we would like to stress that the potential interest for this kind of approach needs not be confined to higher-spin theories \tbl{\cite{SV},\cite{CF}}.

Actually, while our main goal was the elimination of the transversality constraints \eqref{tdiff}, it should be stressed that the mere existence of alternative parametrizations is by no means restricted to theories whose gauge symmetry involves in some formulations constraints of a given sort. Possibly the simplest example is provided by Maxwell's theory itself. Indeed, nothing forbids to ``solve'' for the standard scalar parameter $\L$ in $\d A_{\, \m} = \pr_{\, \m} \L$ in terms of the divergence of a vector parameter $\e^{(0)}_{\, \a}$,
\be
\L \, = \, \pr^{\, \a}\, \e^{(0)}_{\, \a} \, .
\ee
Furthermore, the latter form of the spin$-1$ gauge invariance is clearly reducible, with a chain of gauge-for-gauge transformations essentially identical to that of the spin$-2$ case \eqref{spin2gfg}, up to a shift in the form degree:
\begin{align} \label{spin1gfg}
&\L =  \pr^{\, \a} \e^{(0)}_{\a}& &  \e^{(0)} : && {\small \young(\hfil)}\, ,  && o(\pr^{2})\, , &&\\
&\d \e^{(0)}_{\, \a} =  \pr^{\, \b_1} \e^{(1)}_{[\b_1, \, \a]}& &  \e^{(1)} : && {\small \young(\hfil,\hfil)}\, ,  && o(\pr^{3}) \, ,    && \\
&\cdots& &  &&   &&     && \\
&\d \e^{(k-1)}_{[\b_{k-1},\, \cdots,\, \b_1, \a]} =  \pr^{\, \b_k}\e^{(k)}_{[\b_{k},\,\cdots,\, \b_1, \a]} & &  \e^{(k)} : && {\small \young(0,1,\vdots,k)}\, ,  && o(\pr^{k+2}) \, ,    && \\
\end{align}
with $0 \leq k \leq D -1 $.  To be sure, counting the degrees of freedom  using this parametrization we find
\be
D \, - \, \sum_{k = 0}^{D-1}\, (-1)^k \, (k + 3) \, {D \choose k + 1} \, = \, D - 2\, ,
\ee
for $D > 1$. More generally, one can consider higher-order parametrizations of the form
\be
\L \, = \, \pr^{\, \a_1}\cdots \pr^{\, \a_l} \, \e^{(0)}_{\, \a_1 \, \cdots \, \a_l}\, ,
\ee
together with the corresponding pattern of gauge-for-gauge transformations,
\begin{align} \label{spin1gfg}
&\d \e^{(k-1)}_{\b_{k-1},\, \cdots,\, \b_1, \a_l} =  \pr^{\, \b_k}\e^{(k)}_{\b_{k},\, \cdots,\, \b_1, \a_l} & &  && \e^{(k)} : &&{\small \overbrace{\young(\hfil \hfil \cdots \hfil,1,\vdots,k)}^{l}} ,  && o(\pr^{k+l + 1}) \, ,    && \\
\end{align}
with $0 \leq k \leq D -1$, while the rank $l$ of the symmetric parameter $\e^{(0)}$ is essentially arbitrary. Indeed, taking into account the dimensions of the corresponding tableaux,
\be \label{dimhook}
\mbox{dim}_{\, GL(D)} \, {\small \overbrace{\young(\hfil \hfil \cdots \hfil,1,\vdots,k)}^{l}}  \, =  \, {D + l - 1 \choose l + k } \, {l + k - 1 \choose k} \, ,
\ee
where $l \geq 2$, it is possible to check that the total amount of gauge symmetry is correctly accounted for in this case as well.

 The aforementioned considerations extend to higher spins in a direct way. For instance, the standard vector parameter $\L_{\, \m}$ of the massless Fierz-Pauli theory (no transversality conditions assumed) could be ``solved for'' in terms of the divergence of a rank-$2$ tensor as
\be
\L_{\, \m} \, = \, \pr^{\, \a} \, \l_{\,\a \m}.
\ee
The resulting system of gauge-for-gauge transformations would then be easily recovered using the results of section \ref{sec:traceful}, starting from the solution to the equation $ \pr^{\, \a} \, \l_{\, \a \m} = 0$ in the form \eqref{solve0}\ft{Let us mention, however, that strictly speaking the non-standard forms of the gauge symmetry presented in this section do not properly qualify as {\it resolutions}, in the sense defined in section \ref{sec:syz}.  Indeed, a resolution always gives a local generating set in the space of  gauge symmetries, so that {\it any} gauge  symmetry can be expanded on this set with {\it local} coefficients. Differently, it is in general not possible to represent the standard gauge transformation as a particular form of the non-standard one.}.

Whether or not exploring the space of possible resolutions (or alternative parametrizations) might provide some additional insights into the structure of interacting gauge theories is not clear (and not obvious) to us at this level. However, the mere existence of these alternatives calls for a more complete understanding of their possible role. This last observation applies both to those cases where non-abelian deformations of the standard gauge symmetry are available and, clearly, whenever such deformations are known to be obstructed.

%%%%%%%%%%%%%%%%%%%%%%%%%%%%%%%%%%%%%%%%%%%%%%%%
\section*{Acknowledgments}
%%%%%%%%%%%%%%%%%%%%%%%%%%%%%%%%%%%%%%%%%%%%%%%%

We would like to thank the Galileo Galilei Institute for Theoretical Physics for the hospitality during the workshop ``Higher Spins, Strings and Duality'', and the INFN for partial support during the completion of this work. We are grateful to A. Campoleoni, T. Erler, M. Grigoriev, E.~Skvortsov and M.~Vasiliev for useful discussions. The research of D.F. was supported in part by Scuola Normale Superiore, by Centro E. Fermi, by INFN (I.S. TV12) and by the MIUR-PRIN contract 2009-KHZKRX. The work of S.L. and A.Sh. was partially supported by  the RFBR grant 13-02-00551. A.Sh. appreciates the financial support from Dynasty Foundation, S.L. acknowledges support from the RFBR grant 11-01-00830-a.

%%%%%%%%%%%%%%%%%%%%%%%%%%%%%%%%%%%%%%%%%%%%%%%%
\begin{appendix}
%%%%%%%%%%%%%%%%%%%%%%%%%%%%%%%%%%%%%%%%%%%%%%%%

%%%%
\section{Appendix}
%%%%

%%
\subsection{Generalities on degrees of freedom count} \label{sec:dof}

 The commonly known way to count the physical degrees of freedom in Lagrangian systems goes through the constrained Hamiltonian analysis. For the case where there are no second class constraints and the gauge symmetry generators are irreducible, a general formula for second order field equations was provided long ago in  \cite{HTZ}:
\be \label{DoFir}
  N\, = \,n \, - \, \sum_{k=0}\, (m+1)\, r_m \, .
\ee
Here $N$ is the number of physical degrees of freedom, $n$ counts the number of field components while $r_m$ is the number of gauge generators of $m$-th differential order, i.e. the maximal order of the derivatives acting on the corresponding gauge parameter in $\d \vf$. However, one would like to be able to control the number of degrees of freedom in more general situations, including the case where the equations of motion do not follow from a Lagrangian.

In \cite{KLS}, a simple receipt has been found for an explicitly covariant count of the physical degrees of freedom for general field equations, be they Lagrangian or not. The resulting formula also covers the case of reducible symmetries as well as theories with second-class constraints. For the case of involutive second order Lagrangian equations\footnote{Second order PDE's are termed \emph{involutive} whenever they have no first or zero order differential consequences. The Lagrangian field equations (\ref{Mlagr}) are involutive. Notice that Lagrangian equations are not always involutive. For instance, Proca's equations for the massive vector field $A_{\, \m}$, being of second order, are not involutive as they have the first-order differential consequence $\prd A=0$.} with reducible gauge symmetry the general formula for the number of physical degrees of freedom given in \cite{KLS} reduces to
\be
\label{DoFapp} N \, = \, n \, + \, \sum_{k,m=0}(-1)^{k+1}\,
(m+1)\, r_m^{(k)} \, ,
\ee
where $N$ and $n$ have the same meaning as in \eqref{DoFir}  (with $n$ coinciding  with the number of the second order field equations) while $r_m^{(k)}$ is the number of gauge symmetries of the reducibility order $k$ and of differential order $m$, in the sense specified above. The reducibility order refers to the generation of gauge-for-gauge transformation, so that the gauge parameter appearing in $\d \vf$ is of zero reducibility order, while transformations of this parameter leaving the field invariant involve parameters of first reducibility order and so on.

 In computing the differential order of a parameter one has to take into account all derivatives acting on it in the corresponding gauge transformation for the gauge potential. For instance if the original gauge symmetry is reducible and the first-order reducibility parameter is involved in the transformation with $l_1$ derivatives, then the differential order of the first reducibility order transformation $m_1$ is defined as $m_1= m_0 + l_1$, where $m_0$ is the differential order of the original parameter. A similar prescription was also suggested for gauge fixed Lagrangian systems in \cite{siegel}.

 As a simple illustration of the formula \eqref{DoFapp} let us compute the degrees of freedom for spin$-2$ massless fields in $D$ dimensions, with the chain of gauge-for-gauge transformations \eqref{spin2gfg}
\be
\d\, \e^{(k)}_{[\a,\, \b,\, \m_1,\, \cdots, \, \m_k]} \, = \, \pr^{\, \m_{k+1}}\, \e^{(k+1)}_{[\a,\, \b,\, \m_1,\, \cdots, \, \m_k, \, \m_{k+1}]} \, , \qquad k \, = \, 0 \, ,\cdots, D-2 \, .
\ee
The $(k+2)-$form  $\e^{(k)}$ has ${D \choose k+2}$ components and appears in $\d \vf$ together with $k + 2$ derivatives. Substituting in the  \eqref{DoFapp} we obtain, for the case of traceless $\vf_{\, \m \n}$,
\be \label{DoFs2dd}
N\, =\, \left(\frac{D \, (D+1)}{2} -1\right)  -
\sum_{k=0}^{D-2}\, (-1)^{k} \, (k+3)\, \binom{D}{k+2}\, = \, \fr{D\,
(D - 3)}{2} \, ,
\ee
which is the right number of polarizations for the massless spin$-2$ particle in $D$ dimensions. For traceful fields the only difference would be to add one unit (corresponding to the trace of $\vf_{\, \m \n}$) to \eqref{DoFs2dd}, providing the additional propagating scalar present in that case.

\subsection{Dimensions of irreducible representations of $GL(D)$ and $O(D)$}  \label{sec:dimension}

In this section we collect two useful formulae providing the dimensions of irreps $ V^{\, GL(D)}_{\l}$  of $GL(D)$ and  $V^{\, O(D)}_{\l}$ of $O(D)$. The notation refers to the corresponding Young diagrams, denoted with $\l$.

\subsubsection*{$GL(D)$}
\be \label{GL}
dim \, V^{\, GL(D)}_{\l} \, = \, \prod_{k=1}^N \, \fr{D \, - \, r_k \, + \, c_k}{h_k} \, ,
\ee
where $N$ denotes the total number of boxes, each box is identified with a number from $1$ to $N$, $r_k$ and $c_k$ label row (counted from top to bottom) and column (counted from left to right) of the box labeled with $k$. $h_k$ is the {\it hook length} of the same box, i.e.  the total number of cells forming the hook $\tiny \young(\bullet\rightarrow,\downarrow)$ having the given box as its vertex  \cite{hamermesh}, as indicated in the following example:
\be
\young(96431,7421,41,2,1)
\ee
\subsubsection*{$O(D)$}
\be \label{O}
dim \, V^{\, O(D)}_{\l} \, = \, \fr{1}{h} \, \prod_{i = 1}^n \, \fr{(D  +  s_i  -  n  -  i  -1)!}{(D - 2i)!} \, \prod_{j = i}^n \, (D  +  s_i + s_j -  i  - j)\, ,
\ee
where $h = \prod_{k = 1}^N h_k$ is the hook length of the diagram, $n$ is the total number of rows, $s_i$ denotes the length of the $i$-th row, with rows counted from top to bottom \cite{king}.

\subsection{Spin s in D=4 in spinorial notation}  \label{sec:spinor}

In $D=4$, it might be worthwhile to write the gauge symmetry of \eqref{Mlagr} in terms of two-component spinors. In this formalism rank$-s$ traceless tensors correspond to spinor-tensor fields $\varphi^{\, \alpha_1\cdots\alpha_s\, \dot\alpha_1\cdots\dot\alpha_s}$, totally symmetric in dotted and in undotted spinor indices separately. The number of components of this field is $n=(s+1)^2$. The gauge transformation $\d \vf = \pr \L$ in spinorial form reads
\begin{equation} \label{SGT}
\delta\, \varphi^{\, \alpha_1\cdots\alpha_s\,
\dot\alpha_1\cdots\dot\alpha_s}\, =\,
\partial^{\, (\alpha_1\dot\alpha_1}\, \L^{\, \alpha_2\cdots\alpha_s\, \dot\alpha_2\cdots\dot\alpha_s)} \,.
\end{equation}
Here the round brackets mean symmetrization in dotted and in undotted indices {\it separately}. The resolution of the transversality constraint \eqref{solve} in this notation takes the form
\begin{equation}\label{SGT1}
\L^{\alpha_2\cdots\alpha_s\, \dot\alpha_2\cdots\dot\alpha_s}\, =\,
i\left( \partial^{\, \dot\alpha_2}{}_{\beta_2}\cdots
\partial^{\, \dot\alpha_s}{}_{\beta_s} \varepsilon^{\alpha_2\cdots\alpha_s\, \beta_2\cdots\beta_s}-c.c.\right)\, .
\end{equation}
where the gauge parameter $\varepsilon$ is an arbitrary totally symmetric spin-tensor of rank $2s-2$. It has $4s-2$ (real) components and corresponds to $\e^{(0)}$ in the main body of the paper. The gauge transformation for $\varepsilon$ reads
\begin{equation}\label{SGT2}
\delta \, \varepsilon^{\, \alpha_2\cdots\alpha_s\,
\beta_2\cdots\beta_s}\, =\,
\partial^{\, (\alpha_2}{}_{\dot\beta_2}\cdots
\partial^{\, \alpha_s}{}_{\, \dot\beta_s}\, \omega^{\, \beta_2\cdots\beta_s)\dot\beta_2\cdots\dot\beta_s}\, ,
\end{equation}
where $\omega^{\, \beta_2\cdots\beta_s\, \dot\beta_2\cdots\dot\beta_s}$ has $s^2$ components and corresponds to the totally symmetric traceless tensor of the rank $s-1$ or, upon dualizing in one index, it can be described by the tensor with a hook-type Young tableau, having $s-1$ cells in the row. The transformation of the parameter $\varepsilon$ involves $s-1$ derivatives. As the parameter $\varepsilon$ is itself involved with $s$ derivatives in the original gauge transformation ($k_0=s$), the overall differential order of $\o$ is $2s-1$.

The second-order reducibility transformations look
\begin{equation}\label{SGT3}
\delta \, \omega^{\, \beta_2\cdots\beta_s\, \dot\beta_2\cdots\dot\beta_s}\, =\, \partial^{\, (\beta_2\dot\beta_2}\rho^{\, \beta_3\cdots\beta_s\dot\beta_3\cdots\dot\beta_s)}\, ,
\end{equation}
where $\rho^{\beta_3\cdots\beta_s\dot\beta_3\cdots\dot\beta_s}$ has $(s-1)^2$ components. In tensorial form it corresponds to a totally symmetric traceless tensor of rank $s-2$ or, after dualization, to a hook-type tensor, with $s-2$ cells in the first row as in \eqref{eps3}. Taking into account the differential order of $\o$ and substituting all relevant numbers in \eqref{DoF} one can check that \eqref{SGT1}, \eqref{SGT2} and \eqref{SGT3} correctly describe the degrees of freedom of these systems:
\begin{equation}
N= (s+1)^2-(s+1)(4s-2)+2s\cdot s^2-(2s+1)(s-1)^2=2\,.
\end{equation}
\subsection{Spin 3 in D=3} \label{sec:spin3d3}

 In space-time dimension lower than $4$, parameters that are transverse and traceless  do not admit the parametrization \eqref{solve}, since the corresponding window diagrams do not exist.  In these cases the general solution to the transversality condition \eqref{tdiff} involves more than $s-1$ derivatives, similarly to the case of the $\e^{(D-3)}$ parameters for $D \geq 4$. Considering for instance the case of spin $3$, one finds that a third-order transformation is needed, since in the spirit of our discussion of section \ref{sec:traceless}, we should look for a parametrization for $\L_{\, \m \m}$ involving one more derivative, schematically
\be \Lambda \, \sim \, \partial^2\, \young(\hfil\hfil,\hfil\pr)\, ,
\ee
where the unconstrained parameter is a $\{2, \, 1\}-$traceless tensor or, by duality, a symmetric traceless tensor $\omega$. More explicitly,
\begin{equation}\label{sol}
\L_{\, \m \m}\, =\, \varepsilon_{\, \a \b \m}\, \partial_{\, \m}\,
\partial^{\, \a}\, \partial^{\, \r}\, \o^{\, \b}{}_{\, \r}\
-\varepsilon_{\, \a \b \m}\, \partial^{\, \a}\, \Box \,
\o_{\m}{}^{\b}\,,
\end{equation}
where $\ve_{\, \a \b \g}$ is the Levi-Civita symbol. The gauge-for-gauge transformations for $\o$ in their turn involve a vector parameter $\rho$:
\be \delta\, \o_{\, \m \m} \, =\, \partial_{\, \m}\, \rho_{\, \m} \,
- \, \frac13\, \eta_{\, \m \m} \prd \rho \, . \ee
Counting the physical degrees of freedom we obtain
\be
N \, =\, 7\, - \, 5 \cdot 5\, +\, 6\cdot 3\, =\, 7\, -\, 25\,
+\, 18\,  =\, 0\, ,
\ee
as expected, since no degrees of freedom are associated to massless particles with $s>1$ in $D\leq3$.

%%%%
\end{appendix}
%%%%

%%%%%%%%%%%%%%%%%%%%%%%%%%%%%%%%%%%%%%%%%%%%%%%%
%% BACKMATTER
%%%%%%%%%%%%%%%%%%%%%%%%%%%%%%%%%%%%%%%%%%%%%%%%

%%%%%%%%%%%%%%%%%%%%%%%%%%%%%%%%%%%%%%%%%%%%%%%%
%% REFERENCES
%%%%%%%%%%%%%%%%%%%%%%%%%%%%%%%%%%%%%%%%%%%%%%%%


\begin{thebibliography}{99}

\bibitem{SV}
  E.~D.~Skvortsov and M.~A.~Vasiliev,
  %``Transverse Invariant Higher Spin Fields,''
{\it  Phys.\ Lett.\ B} {\bf 664} 301 (2008) arXiv:hep-th/0701278.
  %%CITATION = HEP-TH/0701278;%%

\bibitem{CF}  A.~Campoleoni and D.~Francia,
%``Maxwell-like Lagrangians for higher spins,''
{\it JHEP} {\bf 1303} (2013) 168 [arXiv:1206.5877 [hep-th]].
%%CITATION = ARXIV:1206.5877;%%

\bibitem{fierz}
M.~Fierz,
%Uber die relativistische Theorie kraftefreier Teilchen mit  beliebigem Spin
{\it Helv.\ Phys.\ Acta}  {\bf 12} (1939) 3.

\bibitem{D07}
D.~Francia
 %``Geometric Lagrangians for massive higher-spin fields,''
{\it Nucl.\ Phys.\  B } {\bf 796}  77 (2008) arXiv:0710.5378 [hep-th].
 %%CITATION = NUPHA,B796,77;%%

\bibitem{fronsdal}
C.~Fronsdal
%``Massless Fields With Integer Spin,''
{\it Phys.\ Rev.\ D} {\bf 18}  3624 (1978).
%%CITATION = PHRVA,D18,3624;%%

\bibitem{PT}
A.~Pashnev and M.~Tsulaia,
%``Description of the higher massless irreducible integer spins in the BRST approach,''
{\it Mod.\ Phys.\ Lett.\ A}  {\bf 13} (1998) 1853 [hep-th/9803207].
%%CITATION = HEP-TH/9803207;%%

\bibitem{fs3}
D.~Francia and A.~Sagnotti,
%``Minimal local Lagrangians for higher-spin geometry,''
{\it Phys.\ Lett.\ B} {\bf 624} (2005) 93 [hep-th/0507144].
%%CITATION = HEP-TH/0507144;%%

\bibitem{Maxim}
  M.~Grigoriev,
  %``Parent formulations, frame-like Lagrangians, and generalized auxiliary fields,''
  JHEP {\bf 1212} (2012) 048
  [arXiv:1204.1793 [hep-th]].
  %%CITATION = ARXIV:1204.1793;%%

\bibitem{fs1}  D.~Francia and A.~Sagnotti,
%``Free geometric equations for higher spins,''
{\it Phys.\ Lett.\ B} {\bf 543} (2002) 303 [hep-th/0207002].
%%CITATION = HEP-TH/0207002;%%

\bibitem{fms1}
D.~Francia, J.~Mourad and A.~Sagnotti,
%``Current Exchanges and Unconstrained Higher Spins,''
{\it Nucl.\ Phys.\ B} {\bf 773} (2007) 203 [hep-th/0701163].
%%CITATION = HEP-TH/0701163;%%

\bibitem{D10}
D.~Francia,
%``String theory triplets and higher-spin curvatures,''
{\it Phys.\ Lett.\ B} {\bf 690} (2010) 90 [arXiv:1001.5003 [hep-th]].
%%CITATION = ARXIV:1001.5003;%%

\bibitem{triplets1}
S.~Ouvry and J.~Stern,
  %``Gauge Fields Of Any Spin And Symmetry,''
{\it Phys.\ Lett.\ B} {\bf 177}  335 (1986).
  %%CITATION = PHLTA,B177,335;%%
A.~K.~H.~Bengtsson,
  %``A Unified Action For Higher Spin Gauge Bosons From Covariant String Theory,''
{\it Phys.\ Lett.\ B} {\bf 182}  321 (1986).
  %%CITATION = PHLTA,B182,321;%%

\bibitem{triplets2}
Henneaux M. and Teitelboim C. 1988 {\it Quantum Mechanics of
Fundamental Systems, 2}, eds. Teitelboim C and Zanelli J (Plenum
Press, New York) p 113

\bibitem{fs2} D.~Francia and A.~Sagnotti
%``On the geometry of higher-spin gauge fields,''
{\it Class.\ Quant.\ Grav.}  {\bf 20} S473 (2003)
arXiv:hep-th/0212185.
  %%CITATION = CQGRD,20,S473;%%

\bibitem{st}
A.~Sagnotti and M.~Tsulaia
%``On higher spins and the tensionless limit of string theory,''
{\it Nucl.\ Phys.\ B} {\bf 682} 83 (2004) arXiv:hep-th/0311257.
%%CITATION = HEP-TH 0311257;%%

\bibitem{BV}
I.~A.~Batalin and G.~A.~Vilkovisky,
%``Quantization of Gauge Theories with Linearly Dependent Generators,''
{\it Phys.\ Rev.\ D} {\bf 28} (1983) 2567  [Erratum-ibid.\ D {\bf 30} (1984) 508].
%%CITATION = PHRVA,D28,2567;%%

\bibitem{KLS}
D.~S.~Kaparulin, S.~L.~Lyakhovich and A.~A.~Sharapov,
%``Consistent interactions and involution,''
{\it JHEP} {\bf 1301} (2013) 097 [arXiv:1210.6821 [hep-th]].
%%CITATION = ARXIV:1210.6821;%%

\bibitem{uni-ham}
W.~G.~Unruh,
%``A Unimodular Theory Of Canonical Quantum Gravity,''
{\it Phys.\ Rev.\ D} {\bf 40} (1989) 1048;
%%CITATION = PHRVA,D40,1048;%%
M.~Henneaux and C.~Teitelboim,
%``The Cosmological Constant And General Covariance,''
{\it Phys.\ Lett.\ B} {\bf 222} (1989)  195.

\bibitem{uni-new}
E.~Alvarez, D.~Blas, J.~Garriga and E.~Verdaguer,
%``Transverse Fierz-Pauli symmetry,''
{\it Nucl.\ Phys.\  B} {\bf 756} (2006) 148 [arXiv:hep-th/0606019];
%%CITATION = NUPHA,B756,148;%%

\bibitem{pope}
C.~N.~Pope and K.~S.~Stelle,
%``SU(infinity), SU+(infinity) AND AREA PRESERVING ALGEBRAS,''
{\it Phys.\ Lett.\ B} {\bf 226} (1989) 257.
%%CITATION = PHLTA,B226,257;%%

\bibitem{AHV}
E.~Alvarez and M.~Herrero-Valea,
%``Unimodular gravity with external sources,''
{\it JCAP} {\bf 1301} (2013) 014
[arXiv:1209.6223 [hep-th]].
%%CITATION = ARXIV:1209.6223;%%

\bibitem{LM} J. M. F. Labastida and T. R. Morris,
%`` Massless mixed-symmetry bosoic free fields,''
{\it Phys. Lett. B} \textbf{180} (1986) 101.

\bibitem{E} D. Eisenbud,  {\it The geometry of syzygies. A second course in commutative algebra and algebraic geometry},
Graduate Texts in Mathematics 229 (New York: Springer-Verlag. xvi+243, 2005)

\bibitem{D-VH}  M. Dubois-Violette and M. Henneaux,
% ``Tensor fields of mixed Young symmetry type and N-complexes,''
{\it Commun. Math. Phys.} {\bf 226} (2002) 393-418.

\bibitem{BB} X. Bekaert and N. Boulanger,
%``Tensor gauge fields in arbitrary representations of GL(D,R) : duality & Poincare lemma,''
{\it Commun. Math. Phys.} {\bf 245} (2004) 27-67.

\bibitem{HTZ}
M.~Henneaux, C.~Teitelboim and J.~Zanelli,
%``Gauge Invariance And Degree Of Freedom Count,''
{\it Nucl.\ Phys.\ B} {\bf 332} (1990) 169.
%%CITATION = NUPHA,B332,169;%%

\bibitem{siegel}
W.~Siegel,
%``Hidden Ghosts,''
{\it Phys.\ Lett.\ B} {\bf 93} (1980) 170.
%%CITATION = PHLTA,B93,170;%%

\bibitem{hamermesh}
M.~Hamermesh, {\it Group theory and its applications to physical
problems} (Dover Publications, New York, NY, USA, 1969)

\bibitem{king}
R.~C.~King,
%�The dimensions of irreducible tensor representations of the orthogonal and symplectic groups,�
{\it Canad. \ Jour. \ Math.} {\bf 23} (1971) 176.
\end{thebibliography}
\end{document}